\documentclass{aastex}
\usepackage{spr-astr-addons}
\shorttitle{Decaying sterile neutrinos}
\begin{document}
\title{Can decaying sterile neutrinos account for all dark matter?}

\author{Man Ho Chan}
\affil{Department of Science and Environmental Studies, The Hong Kong 
Institute of Education \\
Tai Po, New Territories, Hong Kong, China}
\email{chanmh@ied.edu.hk}

\begin{abstract}
The recent discovery of unexplained X-ray line of $3.5-3.6$ keV emitted 
from the Perseus cluster of galaxies and M31 and the excess X-ray line of $8.7$ keV 
emitted from the Milky Way center may indicate that dark matter would decay. 
In this article, I show that approximately 80 \% of dark matter being 7.1 
keV sterile neutrinos and 20 \% of dark matter being 17.4 keV sterile 
neutrinos can satisfactorily 
explain the observed X-ray lines and account for all missing mass. No free 
parameter is needed in this model. This scenario is also compatible with 
current robust observational constraints from the matter power spectrum in 
large-scale structures and would alleviate the challenges faced by the 
existing dark matter models. 
\end{abstract}

\section{Introduction}
The dark matter problem is one of the major problems in astrophysics. It is 
commonly believed that some unknown massive and collisionless particles 
exist to account for the missing mass in our universe. They are regarded 
as cold dark matter (CDM) because they are massive and become 
non-relativistic when they decouple from normal matter. In general, the 
CDM model provides good fits on 
large-scale structure observations such as cluster mass profiles 
\citep{Pointecouteau} and the matter power spectrum \citep{Spergel}. However, 
some observations in dwarf 
galaxies and clusters indicate that cores exist 
\citep{Salucci,Borriello,Oh,Tyson,Sand,deBlok,Newman}, which contradict 
the results from N-body simulations based on the CDM model 
\citep{Navarro}. This discrepancy is 
now known as the core-cusp problem. Moreover, the number of small dark 
halos and the density of subhalos predicted by the CDM model does not 
match the observations of Local 
Group \citep{Boylan,Cho}. These discrepancies are respectively known as 
the missing satellite problem and the too big to fail problem.

Although there are some suggestions that baryonic processes such as 
supernovae and AGN feedbacks may help to alleviate the problems 
\citep{Weinberg,Maccio}, 
it is still quite controversial to make conclusion because the baryonic 
component is just a minor component in galaxies and clusters which may 
not have a significant effect on the dark matter distribution 
\citep{Penarrubia,Vogelsberger}. 

Another idea to solve the above problem is that the dark matter is not 
cold. The existence of keV mass dark matter particles, as a candidate of 
warm dark matter (WDM), has been proposed. In particular, one popular 
candidate of WDM particle is sterile neutrino, which does not interact 
with other particles except by gravity. Basically, the WDM model can solve 
the core-cusp problem, the missing satellite problem and the too big to 
fail problem faced by the CDM model \citep{Destri,Lovell}. 
Moreover, some unexplained X-ray fluxes with energies $(3.55-3.57)\pm 
0.03$ keV 
\citep{Bulbul} and 8.7 keV \citep{Koyama,Prokhorov} are obtained recently, which 
may be due to the decay of 
sterile neutrinos. These sterile neutrinos can be produced via some 
mechanisms such as Dodelson-Widrow (DW) mechanism 
\citep{Dodelson} and Shi-Fuller (SF) mechanism \citep{Shi}. 

However, recent observations of the Lyman-$\alpha$ forest and X-ray background 
put a very tight constraint on sterile neutrino mass $m_s$ and the mixing 
angle $\theta$, which nearly rule out the possibility of DW sterile 
neutrinos as dark matter \citep{Seljak,Viel,Boyarsky,Viel2}. 
Nevertheless, these constraints would be released if we do not assume 
that all sterile neutrino dark matter were produced by non-resonant (DW) 
mechanism. For example, the sterile neutrinos that was produced from Higgs 
decays or in split seesaw mechanism can give agreements with both the 
Lyman-$\alpha$ bounds and X-ray bounds \citep{Petraki,Kusenko}. Moreover, 
\citet{Abazajian} shows that the resonantly-produced (produced by SF 
mechanism) decaying sterile neutrinos can solve the problems faced by the 
CDM model and account for all dark matter. However, this 
model requires one more free parameter, the lepton asymmetry $L$, which is 
not known. In this 
article, I show that if all dark matter consists of two types of sterile 
neutrinos produced by DW mechanism, it can satisfactorily account 
for the unexplained X-ray flux and does not contradict to the matter power 
spectrum constrained by the data of the Lyman-$\alpha$ forest. No free 
parameter is needed in this model.

\section{The decay lines from Perseus and the Milky Way}
Recently, a potential detection of an X-ray line at energy 
$E=(3.55-3.57)\pm 
0.03$ keV from a stacked combination of clusters, with a particularly 
bright signal from Perseus cluster, was found \citep{Bulbul}. The largest 
detected 
line flux from the Perseus cluster is $5.2^{+3.7}_{-2.1} \times 10^{-5}$ 
photons cm$^{-2}$ s$^{-1}$ \citep{Bulbul}. A similar line at nearly the 
same 
energy $E=(3.53 \pm 0.03)$ keV from M31 is also reported 
\citep{Boyarsky2}. The largest 
detected flux from the inner 3.4 kpc is $4.9^{+1.6}_{-1.3} \times 10^{-6}$ 
photons cm$^{-2}$ s$^{-1}$ \citep{Boyarsky2}. These X-ray fluxes seem to be larger than the expected fluxes, which require some new mechanisms to explain. There has been a lot of debate over this issue \citep{Boyarsky4,Bulbul2,Jeltema,Jeltema2,Phillips}. Therefore, this problem has not been settled, and we assume that the production of the large detected fluxes involve some unknown mechanisms. Besides, some emission lines above 
6 keV from Milky Way center has been detected. In particular, \citet{Prokhorov} 
find a significant excess of $E=8.7$ keV photons which cannot be 
explained by ionization and recombination processes. The measured intensity of the FeXXVI Ly$\gamma$ is $1.77^{+0.62}_{-0.56}\times 10^{-5}$ ph cm$^{-2}$ s$^{-1}$, which is about 3 times larger than the expected intensity $6.3^{+0.4}_{-0.4}\times 10^{-6}$ ph cm$^{-2}$ s$^{-1}$ and an order of magnitude larger than the general X-ray intensity ($\sim 10^{-6}$ ph cm$^{-2}$ s$^{-1}$) \citep{Ruchayskiy}. The excess flux is $(1.1 \pm 0.6) \times 10^{-5}$ photons cm$^{-2}$ s$^{-1}$ \citep{Prokhorov}. 

It has been suggested that these unexplained lines may be produced by the 
decay of dark 
matter particles \citep{Prokhorov,Bulbul,Boyarsky2,Finkbeiner}. If these 
lines are produced by the decay of 
sterile neutrinos, the corresponding masses of the sterile neutrinos would 
be 7.1 keV and 17.4 keV respectively (since $E=m_s/2$). There are many proposed models 
to generate more than 1 sterile neutrinos. For example, \citet{Asaka} 
propose 
an extension of the Minimal Standard Model ($\nu$MSM) to generate 3 
sterile right-handed neutrinos. Besides, \citet{Xing} 
propose a type(I+II) seesaw mechanism to construct a $6 \times 6$ flavor 
mixing matrix to generate 3 light or heavy sterile neutrinos. Our 
proposal is basically compatible with the above models. One or more sterile neutrinos can be conjectured in the above mechanisms to account for the observations.

However, some recent studies start to challenge the discovery of the excess 3.55 keV line. For example, \citet{Horiuchi} show that the production of sterile neutrinos via DW mechanism cannot 100\% account for all dark matter. Also, \citet{Anderson} claim that no excess 3.55 keV line is detected from some of the other galaxies and galaxy groups. In particular, observations, including our own galaxy, dwarf spheroidal galaxies and Draco, show negative results for the excess 3.55 keV line \citep{Sorensen,Malyshev,Jeltema3}. \citet{Iakubovskyi} reviews most of the criticisms of the excess 3.55 keV line and concludes that the existing result is still consistent with the decaying dark matter scenario. Moreover, \citet{Ruchayskiy} perform detailed analysis and show that the Draco observation does not exclude the dark matter interpretation of the 3.5 keV line. 

Besides, \citet{Carlson} recently report that the detected 3.55 keV lines in the Milky Way center and Perseus highly 
correlate with the morphology of baryonic matter. This suggests that the 3.55 keV line might be originated from baryonic matter, but not the dark matter. However, this result is expected in the Milky Way because we know that the central mass in many 
galaxies including the Milky Way is dominated by baryonic matter \citep{Lelli,Iocco}. 
Also, many studies have suggested that the total matter distribution in 
galaxies basically follows the baryonic matter distribution 
\citep{Gentile,Lelli}. For the Perseus cluster, the parameters used by 
\citet{Carlson} are questionable. First of all, the mass-concentration 
relation used is significantly different from the one in recent 
literature (see \citet{Schaller}). As a 
result, the concentration parameter used in \citet{Carlson} is 50\% larger 
than the standard one \citep{Sanchez}. This would overestimate the dark matter mass 
for Perseus by a factor of 2 and underestimate the mixing angle 
by a factor of 1.4. Secondly, since Perseus is a cool-cored cluster, the 
central density 
of dark matter would be smaller (see \citet{Chan}). Therefore, the central 
region of Perseus would be dominated by baryonic hot gas and the 3.5 keV 
signal would follow the cool-core morphology. Based on the above arguments, the discovery of the excess 3.55 keV line is 
still a good evidence to support the existence of decaying dark matter.

The mixing angles of 7.1 keV sterile neutrinos calculated by 
the fluxes from Perseus and M31 are $\sin^22 \theta=5.5^{+3.9}_{-2.3} 
\times 10^{-10}$ \citep{Bulbul} and $\sin^22 \theta=(0.2-2) \times 
10^{-10}$ \citep{Boyarsky2}
respectively. For 17.4 keV sterile neutrinos, the mixing angle calculated 
is $\sin^22 \theta=(4.1 \pm 2.2) \times 10^{-12}$ \citep{Prokhorov}. 
Obviously, there is some tension between the calculated mixing angles 
obtained from the data of Perseus and M31. However, the calculations of the mixing 
angles in Perseus and M31 assume that the dark matter density follows the 
NFW profile \citep{Bulbul,Boyarsky2}. The NFW profile generally gives good 
fits to most clusters (including Perseus), but not for most galaxies 
\citep{Grillo,Chae}. It 
has been shown that the NFW profile does not give a good fit to M31 dark 
matter density profile \citep{Banerjee,Chemin}. If a dark matter core 
exists in M31, the resulting mixing angle calculated would be larger and 
the tension would be alleviated. For the DW mechanism, the dark matter velocity distribution is thermal so that the sterile neutrino mass determines whether the density profile is cored or cuspy, provided that the baryonic effects are not important. \citet{Vega} show that 7 keV sterile neutrinos can give the correct abundance of substructures for $1-100$ kpc scales, which disagrees with the NFW profile. Therefore, the mixing angle calculated in \citet{Boyarsky2} may be underestimated. In this article, we only use the mixing angle calculated from the data of Perseus because the obtained density profile is more reliable.

Suppose that the sterile neutrinos were produced by the DW mechanism. The 
relation 
between the mixing angle and the fraction of total mass of sterile 
neutrinos to total dark matter mass $f_s$ is given by 
\citep{Dolgov,Mapelli} \begin{equation}
f_s=0.114 \left( \frac{\sin^22 \theta}{10^{-10}} \right) \left( 
\frac{m_s}{7.1~\rm keV} \right)^2.
\end{equation}
If we assume $m_s=7.1$ keV, by using the mixing angle derived in 
\citet{Bulbul}, we get $f_s=0.60-1$ (note that $f_s \le 1$). 

For the Milky Way, the line absorption due to the gas near 
the Milky Way center is significant (the absorption in Perseus is 
negligible). The mixing angle obtained in 
\citet{Prokhorov} is directly calculated from the observed line intensity 
in \citet{Koyama} without absorption corrections. Therefore, we should 
first re-calculate the mixing angles by considering the optical depth of 
the X-ray photons. The optical depth of the X-ray photons emitted at $r$ is given by
\begin{equation}
\tau(r)=\int_r^{R_g}n_g(r) \sigma dr,
\end{equation}
where $\sigma$ is the absorption cross section, $n_g(r)$ and $R_g$ are the 
number density profile and the total size of gas halo respectively. The 
absorption cross section of a particular ion or atom is given by 
\citep{Daltabuit}
\begin{equation}
\sigma_i= \sigma_{th} \left[ \alpha \left( \frac{E_{th}}{E} 
\right)^s+(1-\alpha) \left( \frac{E_{th}}{E} \right)^{s+1} \right],
\end{equation}
where $\sigma_{th}$, $E_{th}$ and $s$ are parameters for different 
elements. Here, we consider some major absorption atoms and metal ions 
such as Hydrogen, Helium, Carbon, Nitrogen, Oxygen, Silicon, Neon, Sulphur 
and Iron. Therefore, the effective cross section is $\sigma= \sum_ia_i 
\sigma_i$, where $a_i$ is the fraction of the element in the gas. The 
value of $a_i$ can be determined by the metallicity of the gas. 

Near the Milky Way center, the 
number densities of atomic and molecular gases are 0.9 cm$^{-3}$ and 74 
cm$^{-3}$ respectively \citep{Ferriere}, and the metallicity of the gas is 
about $(1-3) Z_{\odot}$ \citep{Muno,Sakano}, which largely enhance the 
absorption. Assuming a constant density profile, the optical depth 
is about 0.21. By eliminating the effect of absorption and taking 
$m_s=17.4$ keV, we get $\sin^22 
\theta=(5.1 \pm 2.7) \times 10^{-12}f_s^{-1}$ and $f_s=0.13-0.23$. In fact, observations indicate that the inner density profile follows $r^{-1.8}$ \citep{Schodel} and the density profile in the outer region is nearly a constant \citep{Ferriere}. Since most of the absorption occurs near the inner region, our assumption using the constant density profile would underestimate the effect of absorption. Hence, the value of $f_s$ should be somewhat larger. Nevertheless, we still use the above calculated range as a conservative estimation. 

If there are two types of decaying sterile neutrinos, one being 7.1 keV 
and the other being 17.4 keV, the fraction of sterile neutrinos would be 
$f_s=0.73-1$. In other words, 
the X-ray line fluxes show that these two types of decaying sterile 
neutrinos can account for all or at least a major part of the dark matter. Nevertheless, the emission lines do not exclude a third type of sterile neutrino dark matter as the lower limit of $f_s$ is well below 1.  

\citet{Bulbul} state that 
the sterile neutrinos produced by the DW mechanism can only account for about 1\% of the dark matter. 
It is because they use the mixing angle derived from a stacked spectrum of 
galaxy clusters. In particular, they neglect the result of Perseus because 
they believe that the 
observed intensity might originate from the cool core of Perseus, but 
not from sterile neutrinos. Nevertheless, \citet{Boyarsky2} discover 
that the outskirts of Perseus cluster also show a similar strong 3.5 keV 
line \citep{Iakubovskyi}. In fact, the results of Perseus are more 
reliable because it is a large, nearby, and well-studied cluster. However, 
the full sample used in \citet{Bulbul} contains many small and 
distant clusters (many clusters with redshift $z>0.15$). The stacked 
analysis 
in \citet{Bulbul} uses a single equation and some scaling laws to model 
the amount of dark matter, which may have large uncertainties. For 
example, the mass of cluster hot gas used in \citet{Bulbul} follows the 
results in \citet{Vikhlinin}. This gas mass fraction is 50\% lower than 
the observed one for high-redshift clusters ($z=0.15-0.3$) \citep{Landry}. 
Therefore, the dark matter content for high-redshift clusters in 
\citet{Bulbul} is overestimated. As a result, a smaller mixing angle is 
obtained because the sample contains many clusters with $z>0.15$. In 
addition, the concentration parameter $c_{500}=3$ used in \citet{Bulbul} 
is significantly deviated from the observed one for high-redshift clusters 
($c_{500}=2.5$) \citep{Niikura}. This would significantly 
overestimate the dark matter mass for high-redshift clusters and give an 
overall smaller mixing angle. In fact, the uncertainties of the empirical scaling relations in cosmological simulations are not small, and the scaling relations used in \citet{Bulbul} are somewhat different from that obtained in recent studies \citep{Mantz}. Therefore, using the result of Perseus is 
more appropriate in our analysis.

\section{Reconciling the claim with other observations}
Observations from the Lyman-$\alpha$ forest and the X-ray background give tight 
constraints on sterile neutrino mass. For example, \citet{Seljak,Viel} 
based on 3000 SDSS data obtained $m_s \ge 15$ keV and 
$m_s \ge 11$ keV respectively. Recently, \citet{Viel2} used the data from 
28 high redshift quasars and got $m_s \ge 21.8$ keV. However, as 
pointed out by 
\citet{Boyarsky}, the evaluations of WDM model based on the 
Lyman-$\alpha$ forest suffer from some major problems, such 
as observational uncertainties, theoretical 
uncertainties related to numerical simulations, and astrophysical 
uncertainties relevant for Lyman-$\alpha$ physics (thermal history 
and ionization history of the intergalactic medium). Moreover, it 
has been shown that the late-time velocity dispersion of dark matter 
particles has a 
significant effect on Lyman-$\alpha$ clouds during structure formation, 
which may give a wrong estimation of the lower bound of $m_s$ 
\citep{Valageas}. These factors are a great challenge for the results 
obtained from the Lyman-$\alpha$ forest data alone.

By combining the WMAP 5 data, two different Lyman-$\alpha$ sets and SDSS 
data, \citet{Boyarsky} get a robust constraint on sterile neutrino 
mass: $m_s \ge 8$ keV (frequentist 99.7 \% confidence) and $m_s \ge 
12.1$ keV (Bayesian 95 \% credible interval). However, 
these bounds 
are obtained by assuming there is only one type of sterile neutrinos. 

If there are two types of sterile neutrinos and they contribute to the 
power spectrum individually, the combined power spectrum is given by 
\citep{Ma}
\begin{equation}
P(k)=\{\frac{\Omega_{s1}}{\Omega_{dm}}
[P_{s1}(k)]^{1/2}+\frac{\Omega_{s2}}{\Omega_{dm}}[P_{s2}(k)]^{1/2} \}^2,
\end{equation}
where $\Omega_{dm}$ is the cosmological density parameter 
for dark matter, $\Omega_{s1}$, $\Omega_{s2}$, $P_{s1}(k)$ and $P_{s2}(k)$ 
are the corresponding cosmological density 
parameters and power spectrums of 7.1 keV and 17.4 keV sterile neutrinos 
respectively. It is more convenient to define a transfer function $T(k)$ 
such that $T(k)=[P_{WDM}(k)/P_{CDM}(k)]^{1/2}$, where $P_{WDM}$ and 
$P_{CDM}$ are the power spectrums for the WDM and CDM models 
respectively. The transfer function can be well described by the 
following analytic form \citep{Destri2}:
\begin{equation}
T(k)= \left[ 1+(k/\kappa)^a \right]^{-b/2},
\end{equation}
where $a=2.304$, $b=4.478$, $\kappa=6.06(m_s/{\rm keV})^{0.84} h$ 
Mpc$^{-1}$. Although this analytic form is derived for one-component dark matter, the velocity distributions of these two types of sterile neutrinos are thermal, and the cross sections are so small such that they nearly do not interact with each other. Hence, the resultant contribution of the two components can be thought of as an effective single-component contribution. Assuming $\Omega_{s1}=0.8 \Omega_{dm}$ and 
$\Omega_{s2}=0.2 \Omega_{dm}$, we get 
$k_{1/2}=11.2$ Mpc$^{-1}$, where $[T(k_{1/2})]^2=1/2$. The robust 
constraints from \citet{Boyarsky} give $k_{1/2} \ge 7.44$ 
Mpc$^{-1}$ (frequentist 99.7 \% confidence) and $k_{1/2} \ge 10.5$ 
Mpc$^{-1}$ (Bayesian 95 \% credible interval). Therefore, the 
resultant $k_{1/2}$ calculated satisfies the current lower bounds from the 
combination results of WMAP 5, Lyman-$\alpha$ sets and SDSS data (see 
Fig.~1). Note that the parameter $k_{1/2}$ used here is different 
from another free-streaming parameter $k_{fs}$ which represents 
the comoving scale of collisionless particles that interact only 
gravitationally. Basically, these two parameters serve different purposes, 
and the parameter $k_{1/2}$ is more convenient to use here because it can 
be obtained directly from the transfer function.

\begin{figure*}
 \includegraphics[width=80mm]{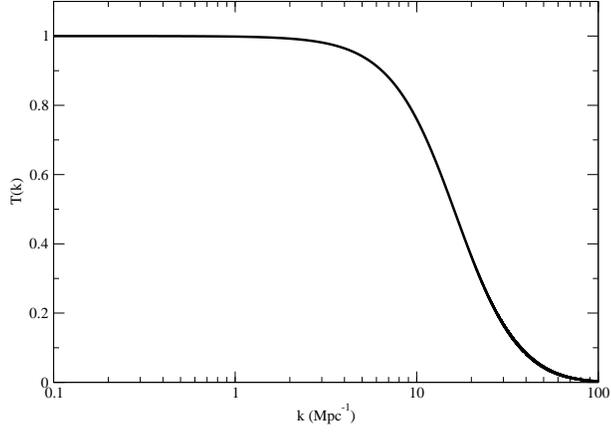}
 \caption{The transfer function $T(k)$ of our model. We assume $h=0.7$.} 
\end{figure*}
\vskip 15mm

The observational data from the model-independent diffuse X-ray background give the following 
constraint \citep{Boyarsky3}:
\begin{equation}
\Omega_s \sin^22 \theta \le 3 \times 10^{-5} \left( \frac{m_s}{\rm keV} 
\right)^{-5}.
\end{equation}
If there are two types of decaying sterile neutrinos, by using Eq.~(1), 
the constraint becomes
\begin{equation}
\Omega_{s1}^2 \left(\frac{m_{s1}}{\rm keV} \right)^3+\Omega_{s2}^2 
\left( \frac{m_{s2}}{\rm keV} \right)^3 \le 2260 \Omega_{dm}^2.
\end{equation}

In Fig.~2, we plot the constraint from diffuse X-ray background together with the 
constraints from decay lines and data from large-scale structures. It 
shows that our results $\Omega_{s1} \approx 0.8 \Omega_{dm}$ with 
$m_{s1}=7.1$ keV and $\Omega_{s2} \approx 0.2 \Omega_{dm}$ with 
$m_{s2}=17.4$ keV satisfy all current constraints. However, as 
mentioned in \citet{Boyarsky}, the single-component non-resonant (DW) 
decaying sterile neutrino model is ruled out by observations (see 
Fig.~2). 

Besides the diffuse X-ray constraints from \citet{Boyarsky3}, our model is also consistent with some recent model-independent constraints. For example, observations of Ursa Minor give model-independent constraints $f_{s1} \le 1.4$ and $f_{s2} \le 0.82$ \citep{Loewenstein}. However, some model-dependent observations give more stringent constraints on $f_{s1}$ and $f_{s2}$. For example, based on the results in \citet{Boyarsky6,Boyarsky5,Sorensen2}, the strongest constraints are $f_{s1} \le 0.5$ and $f_{s2} \le 0.3$. Moreover, some of the model-dependent constraints are completely negative to our decaying sterile neutrino model \citep{Malyshev,Jeltema3}. Nevertheless, most of the results highly depend on the dark matter density profiles of the dwarf spheroidal galaxies, which have large uncertainties. For example, some of the studies follow the `favoured NFW model' \citep{Malyshev,Jeltema3} while recent observations suggest that the dark matter density of the dwarf spheroidal galaxies are cored \citep{Burkert}. The cored profile will suppress the observed flux so that the constraints can be severely alleviated. Besides the functional form of the dark matter density profiles, the parameters of the dark matter density profiles also have large uncertainties, even for our Milky Way and M31. Therefore, these model-dependent constraints cannot completely rule out the possibility of the two-component decaying sterile neutrino model.

\begin{figure*}
 \includegraphics[width=80mm]{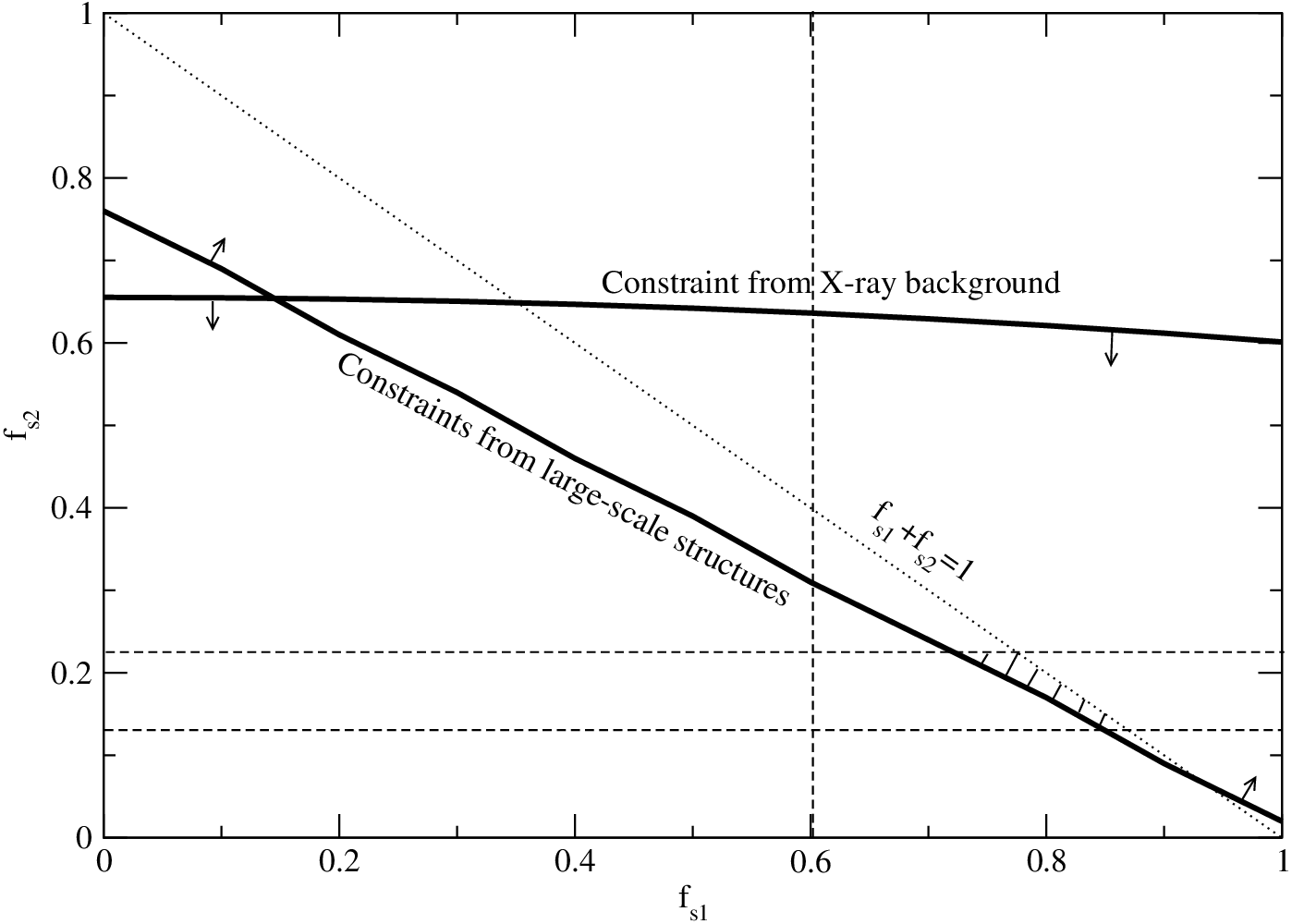}
 \caption{The two solid lines indicate the constraints from large-scale 
structures (WMAP 5 data, two different Lyman-$\alpha$ sets and SDSS data) 
\citep{Boyarsky} and X-ray background \citep{Boyarsky3} respectively. The 
vertical and horizontal dashed lines represent the constraints from decay 
lines ($0.60 \le f_{s1} 
\le 1$ and $0.13 \le f_{s2} \le 0.23$). Here, we define
$f_{s1}=\Omega_{s1}/\Omega_{dm}$ and $f_{s2}=\Omega_{s2}/\Omega_{dm}$. The 
shaded region is the allowed parameter space that satisfies all the 
constraints, including the requirement $f_{s1}+f_{s2} \le 1$ (the dotted 
line).} 
\end{figure*}

\section{Discussion}
In this article, I assume that there are two types of decaying sterile 
neutrinos produced by DW mechanism, which can account for all dark matter 
in our universe without any free parameters. Approximately 80 \% of dark 
matter being 7.1 keV sterile neutrinos 
and 20 \% of dark matter being 17.4 keV sterile neutrinos can 
satisfactorily explain the observed X-ray lines from the Perseus galaxy cluster and the 
Milky Way center. Originally, it has been thought that the decaying 
sterile neutrinos produced by DW mechanism has been ruled out by 
Lyman-$\alpha$ forest data. However, if we include one more type of 
sterile neutrinos, the 
tension in the mass range of sterile neutrinos between the constraint 
from X-ray background and the observed data from power spectrum would be 
released. In general, similar calculations can be applied for decaying 
sterile neutrinos produced by SF mechanism. However, one more free 
parameter would be generated to achieve the same purpose 
\citep{Abazajian}.

Also, we may expect to discover a 8.7 keV line emitted from other 
nearby dwarf galaxies and galaxy clusters. For example, the decaying 
sterile neutrinos in Fornax cluster would give a flux of $\sim 10^{-5}$ 
cm$^{-2}$ s$^{-1}$ or $10^{-13}$ erg cm$^{-2}$ s$^{-1}$ at 8.7 keV. Such a weak line is consistent with the current constraint obtained in \citet{Sorensen}. Further verification can be done by future X-ray observations.

On the other hand, the calculation of the power spectrum in this model 
(Eq.~(4)) does not include the gravitational interaction between the two 
components. A larger $k_{1/2}$ would be obtained if we fully consider 
this effect in the structure formation. Nevertheless, this larger 
$k_{1/2}$ must also satisfy the current constraints from observations. 

If our model is correct, this would alleviate the challenges faced by 
the CDM model (the core-cusp problem, the missing satellite 
problem and the too big to fail problem) and save the 
existing one-component DW sterile neutrino model, which is nearly ruled 
out by the data from large-scale observations. To verify our result based 
on the above calculations, numerical simulations with two-component WDM is 
needed.

\bibliographystyle{spr-mp-nameyear-cnd}
\bibliography{biblio-u1}

\clearpage

\end{document}